\documentclass[12pt]{iopart}
\usepackage{epsfig}
\usepackage{graphicx}

\newcommand{\be}{\begin{equation}}
\newcommand{\ee}{\end{equation}}
\newcommand{\bea}{\begin{eqnarray}}
\newcommand{\eea}{\end{eqnarray}}
\newcommand{\bd}{\begin{displaymath}}
\newcommand{\ed}{\end{displaymath}}

\newcommand{\bra}{\langle}
\newcommand{\ket}{\rangle}

\newcommand{\PD}{{\rm PD}}
\newcommand{\LGD}{{\rm LGD}}

\usepackage{epsfig}
\usepackage{graphicx}
\usepackage{eufrak}

\begin{document}

\title{Credit contagion and credit risk}
\author{JPL Hatchett$^\dag$ and R K\"uhn$^\S$}
\address{$\dag$~ Hymans Robertson LLP, One London Wall, London EC2Y 5EA, UK }
\address{$\S$~ Department of Mathematics, King's College London, The Strand,
London WC2R 2LS, United Kingdom}

\begin{abstract}
We study a simple, solvable model that allows us to investigate
effects of credit contagion on the default probability of individual
firms, in both portfolios of firms and on an economy wide scale.
While the effect of interactions may be small in typical (most
probable) scenarios they are magnified, due to feedback, by
situations of economic stress, which in turn leads to fatter tails
in loss distributions of large loan portfolios.
\end{abstract}


\ead{ \mailto{jon.hatchett@hymans.co.uk}
\mailto{reimer.kuehn@klc.ac.uk}}

\section{Introduction}
Modelling credit risk in a coherent yet applicable manner is an
important yet challenging problem. The difficulties arise from the
combination of a large, and co-dependent set of risk parameters such
as default rates, recovery rates, or exposures, which are correlated
and non-stationary in time. An additional issue is that of credit
contagion \cite{JarrowYu01, GieseckeWeber06, DavisLo01}, which
examines the role of counter-party risk in credit risk modelling. If
a firm is in economic distress, or defaults, this will implications
for any firm which is economically influenced by this given firm,
for example, a service provider to it, purchaser of its goods or a
bank with a credit line to the firm.  The direct correlations
between firms caused by credit contagion lead to further
complications in modelling the overall, either portfolio or economy
wide, level of risk. Jarrow and Yu \cite{JarrowYu01} introduced a
framework of primary and secondary firms, the former would default
depending on some background stochastic process while the latter
were affected by a stochastic process and the performance of the
primary firms. They argued that this was a reasonable level of
detail for their purposes and it also simplifies matters as there
are no feedback loops in the system. Secondary firms depend only on
primary firms whose performance is independent of the secondary
firms. Another approach for modelling credit contagion dynamics was
provided by Giesecke and Weber \cite{GieseckeWeber06} who used the
well known voter process \cite{Liggett99}, from the theory of
interacting particle systems, to model interactions between firms.
They assumed a regular structure for their firms (a regular
hyper-cubic lattice) and focussed on the equilibrium properties of
their model. Davis and Lo \cite{DavisLo01} considered a model in
which defaults occur either directly, or through infection by
another defaulted firm, with probabilities for direct default or
infection taken uniform throughout the system. Defaults occurring
due to both, endogenous or exogenous causes were not considered in
their set-up.

There are a variety of techniques for modelling the correlations
between firms' default behaviour, which is a major complication in
credit risk modelling. The binomial expansion technique assumes
independence between firms so that the number of defaults in a
portfolio is described by a binomial distribution. In order to
capture the effects of correlations a binomial distribution with an
``effective" number of firms is assumed which is {\em smaller\/}
than the actual number in the portfolio, but the weight given to
each firm scaled so as to keep the mean number of defaults constant,
while the variance of the overall number of defaults is increased.
The relationship between the true number of firms and the effective
reduced number is a modelling choice that depends on the diversity
of the firms in terms of sectors, geographic locations or any other
identifiable trait that would lead to strong correlations in default
behaviour. JP Morgans' CreditMetrics approach \cite{Creditmetrics97}
and Credit Suisse First Financial Products CreditRisk$^{+}$
\cite{CreditRiskplus} (see \cite{Gordy00} for a detailed comparison
between the two) uses the correlations in equity values as a
surrogate for the correlations in credit quality. The structural
modelling approach goes back a long way to work by Merton
\cite{Merton73}  which directly models the dynamics of a firm's
assets, with default being triggered by the asset value hitting some
predetermined value (which henceforth we take without loss of
generality to be zero). Correlations between firms are due to
correlations in the dynamics of different firms' assets. This
approach is very general, as it is relatively transparent to
identify different driving forces of asset levels and
straightforward to include them in the model (though the resulting
model itself will be non-trivial). However, it suffers from the fact
that the asset level is not an observable quantity \cite{Siuetal05}.
On the other hand, the reduced form approach gives default rates for
a given firm without modelling the underlying default process.
Correlations are then directly introduced between the default rates.
There was some discussion in the literature about whether the
reduced form model could describe the true level of default
correlations seen empirically but Yu \cite{Yu05} seems to have
answered this question in the affirmative if a suitable structure
between the default rates is taken into account.

The approach we take here is a discrete time Markov process (at the
microeconomic level) where the probability of a default of a given
firm in a particular time step depends materially on the state of
its economic partners at the start of that time step, as well as on
macro-economic influences. Using techniques developed in the
statistical mechanics of disordered systems, and recently applied to
this specific model in \cite{HatchettKuehn06}, we are able to solve
our model exactly, and given our assumptions that we describe
shortly, this solution takes a particularly simple form despite the
fact that in principle we have feedback correlations,
non-equilibrium dynamics and in principle non-Markovian behaviour at
the macroscopic (economy/portfolio wide) level. We note that it is
possible to frame our model in either the structural approach or the
reduced form approach, depending on requirements and taste, although
the interpretation of the variables in the two approaches will of
course be different. We find that the correlations introduced
through credit contagion lead to large increases in default rates in
times of economic stress, above and beyond those introduced by
simple macro-economic dependencies. This has strong implications for
portfolio risk management.

\section{The microeconomic framework}
We will analyse an economy of $N$ firms in the large $N$ limit.
Generally, we focus on the characteristic changes in the economy due
to interactions between firms, which will be described in a probabilistic
manner.

As mentioned in the introduction we take a discrete time approach.
For clarity we restrict our discussion to a one year time frame
split into twelve steps; this is not essential, but parameters may
need rescaling depending on the set-up. We use a binary indicator
variable $n_{i,t}$ to denote whether firm $i$ is solvent at time $t$
($n_{i,t} = 0$) or has defaulted ($n_{i,t} = 1$).  The default
process is a function of an underlying stochastic process for each
firm in terms of a ``wealth'' variable $W_{i,t}$, where we assume
default if the wealth drops below zero. We shall assume that recovery
from default over the time horizon of a year is not possible, so that
the defaulted state is absorbing. As a function of the wealth,
therefore, the indicator variables evolve according to
\be
n_{i,t+1} = n_{i,t} + (1 - n_{i,t}) \Theta(-W_{i,t}) \ ,
\label{eq:indicator_dynamics}
\ee
where $\Theta(\ldots)$ is the Heavyside function.

A {\em dynamic model\/} for the indicator variables is obtained from
(\ref{eq:indicator_dynamics}) by specifying the underlying
stochastic process for the wealth variables $W_{i,t}$. We shall take
it to be of the form \be W_{i,t} = \vartheta_i - \sum_{j = 1}^N
J_{ij} n_{j,t} - \eta_{i,t}\ . \label{eq:wealth_gen} \ee Here
$\vartheta_i$ denotes an {\em initial wealth\/} of firm $i$ at the
beginning of the risk horizon, and $J_{ij}$ quantifies the material
impact on the wealth of firm $i$ that would be caused by a default
of firm $j$. This may or may not be a reduction in wealth, depending
on whether $j$ has a cooperative ($J_{ij}>0$) or a competitive
($J_{ij}<0$) economic relation with $i$.

We shall assume that the fluctuating contributions $\eta_{i,t}$ to
(\ref{eq:wealth_gen}) are zero-mean Gaussians. There is still some
degree of flexibility concerning  the decomposition of the
$\eta_{i,t}$ into contributions that are {\em intrinsic\/} to the
firm and {\em extrinsic\/} contributions. The latter describe the
influence of economy-wide fluctuations or fluctuations pertaining to
different economic sectors, depending on the level of detail
required. We restrict ourselves to a minimal model containing a {\em
single\/} macro-economic factor (assumed to be constant over a risk
horizon of a year), and individual fluctuations for each firm, \be
 \eta_{i,t} = \sigma_i\left(\sqrt{\rho_i} \eta_{0} + \sqrt{1 - \rho_i} ~
\xi_{i,t} \right)\ , \label{eq:noise_dec_gen} \ee where $\sigma_i$
sets the scale of the individual fluctuations, and the
$\{\xi_{i,t}\}$ are taken to be {\em independent\/}
$\mathcal{N}(0,1)$ Gaussians; finally, the parameters $\rho_i$
quantify the correlations of the $\eta_{i,t}$ created via the
coupling to economy-wide fluctuations $\eta_{0}$, also taken to be
$\mathcal{N}(0,1)$.

Up to this point the wealth dynamics does not contain an endogenous
drift. If predictions are required over longer time periods then it
may also be pertinent to introduce such a drift, e.g. by using a
time-dependent $\vartheta_i$ for example, $\vartheta_{i,t} =
\vartheta_i(0) {\rm e}^{z_i t}$, where $z_i$ denotes an intrinsic
growth rate of the average wealth of firm $i$ (with $z_i>0$ for a
firm making profits and $z_i<0$ for a firm making losses). However,
for the current purposes of examining default rates over the medium
term and especially focussing on the behaviour on the tails, this
adjustment does not lead to significant changes in our overall
conclusion.

The model, as formulated above, clearly takes a {\em structural point of
view\/} on the problem of credit contagion. However, we note that the
dynamics (\ref{eq:indicator_dynamics}) of the indicator variables is
clearly independent of the {\em scale\/} of the wealth variables
$W_{i,t}$. By appropriately rescaling the initial wealths $ \vartheta_i$ and
the impact parameters $J_{ij}$ we can thus assume a unit-scale $\sigma_i\equiv
1$ for the noise variables (\ref{eq:noise_dec_gen}).  Interestingly, this
simple rescaling, which leaves the dynamics of the system unaffected,
amounts to changing to a {\em reduced-form\/} interpretation of the
dynamics.

To see this, note from (\ref{eq:wealth_gen}) that the event $W_{i,t} <
0$ is equivalent to $\eta_{i,t} > \vartheta_i - \sum_{j =1}^N J_{ij}
n_{j,t}$.  With $\sigma_i\equiv 1$, we see  that this occurs
with probability $\Phi(\sum_j J_{ij} n_{j,t} - \vartheta_i)$ where
$\Phi(\cdot)$ is the cumulative normal distribution. From a reduced form
point of view this is just the intensity of default of firm $i$ at time step
$t$ (in a given economic environment specified by the set of firms defaulted
at time $t$). This allows us to re-interpret the (rescaled) initial wealth
and impact variables $\vartheta_i$ and $J_{ij}$ in terms of the bare default
probabilities \cite{HatchettKuehn06, KuehnNeu03, NeuKuehn04}.
I.e., if company $i$ has an expected default probability of $p_i$ in a
given time unit (e.g. one month in the present set-up) as predicted from
tables from ratings agencies, then $\vartheta_i = - \Phi^{-1}(p_i)$.
Similarly, the expected default probability $p_{i|j}$ of firm $i$, given
that only firm $j$ has defaulted leads to the value $J_{ij} = \Phi^{-1}
(p_{i|j}) - \Phi^{-1}(p_i)$.

In determining the model parameters by the method suggested above we are
splitting our default probability into terms that come from credit contagion
and other terms such as the bare default probability that come from historical
data. It could fairly be argued that the historical data already incorporate
the credit contagion terms and thus we are double counting. As we will see
later in numerical simulations, the {\em credit contagion terms make very
little difference to average behaviour} and thus making estimates based on
average historical data is still a reasonable approach.

In choosing the variable $\rho_i$ we follow the prescription given
by BASEL II \cite{BASELII} which sets
\begin{eqnarray}
\rho_i = 0.12 \frac{1- \rme^{-50 \PD_i}}{1 - \rme^{-50}}
+ 0.24 \left(1 - \frac{1 - \rme^{-50 \PD_i}}{1 - \rme^{-50}}\right)
\approx 0.12\left(1 + \rme^{-50 \PD_i}\right)
\label{eq:Basel_corr}
\end{eqnarray}
where $\PD_i$ gives the probability of default of firm $i$ over one
year, ignoring credit contagion effects. With $p_i=
\Phi(-\vartheta_i)$ as the monthly default probability, we have
$\PD_i \approx 12 \Phi(-\vartheta_i)$.

We still have to specify the form for the economic interactions. We
adopt here a probabilistic approach, so take them to be random
quantities of the form
\begin{eqnarray}
J_{ij} = c_{ij}\left[\frac{J_0}{c} + \frac{J}{\sqrt{c}} x_{ij} \right]\ .
\label{eq:couplings}
\end{eqnarray}
Here, the $c_{ij} \in \{0,1\}$ detail the network (presence or absence) of
interactions between different firms and we choose these to be randomly
fixed according to
\begin{eqnarray}
P(c_{ij}) = \frac{c}{N} \delta_{c_{ij},1} + \left(1 - \frac{c}{N}
\right) \delta_{c_{ij},0}\ , \qquad i < j\ , \qquad c_{ji} = c_{ij}\ .
\end{eqnarray}
We assume that the average connectivity $c$ of each firm is large in the
limit of a large economy; this will allow the influence of partner firms
to be described by the central limit theorem and the law of large numbers.
Concerning the values of the (non-zero) impact parameters, we parametrise
them as shown, with $x_{ij}$ assumed to be zero-mean, unit-variance random
variables, with finite moments, and {\em independent in pairs},
\begin{eqnarray}
\overline{x_{ij}} = 0\ , \qquad \overline{x_{ij}^2} = 1\ , \qquad
\overline{x_{ij} x_{ji}}= \alpha\ , \qquad \overline{x_{ij} x_{kl}} = 0
~~\mbox{otherwise\ .}
\label{eq:xij_stats}
\end{eqnarray}
The parameters $J_0$ and $J$ determine mean and variance of the interaction
strengths; the scaling of mean and variance with $c$ and $\sqrt c$ respectively
in (\ref{eq:couplings}) is necessary to allow a meaningful large $c$ limit to
be taken. Taking $J_0>0$ would encode the fact that on average firms have a
synergy with their economic partners.

At first sight, specifying the $J_{ij}$ appears to introduce a vast number
of parameters into our model, but in fact only the first two moments of the
distribution of interaction strengths are sufficient to determine the
macroscopic behaviour of the system, and so the model space is not too large.

Let us now turn to the capital required to be held against credit risk. In
the BASEL II document \cite{BASELII} the capital requirement for a unit-size
loan given to firm $i$ is
\begin{eqnarray}
K_i =\LGD_i \left[ \Phi\left(\frac{\sqrt{\rho_i}\,\Phi^{-1}(0.999) +
\Phi^{-1}(\PD_i)}{\sqrt{1 - \rho_i}}\right) - \PD_i\right ] M_i\ .
\label{eq:BaselII_cap}
\end{eqnarray}
The first factor, the loss given default $\LGD_i$ of firm $i$, is
related to the average fraction of a loan that can be recovered
despite default. The last factor, $M_i$, is related to the maturity
(long dated loans are inherently riskier). Adjustments related to
liquidity (low liquidity loans are risker) and concentration (fewer,
larger loans give a greater variance in returns for given expected
return) are occasionally also included in this factor ---
concentration-adjustments, in fact, are a means to account for
reduced granularity in a credit portfolio resulting from the
possibility of credit contagion.

The factor inside square brackets in (\ref{eq:BaselII_cap}) is
entirely related to the loss-frequency distribution. The first term
is the value of the loss frequency not exceeded with probability
$q=0.999$ under fluctuating macro-economic conditions, with $\rho_i$
describing the dependence of the firm's loss-frequency on the
macro-economic factor. The second term is the average loss
frequency. The value of the confidence level $q$ is in principle
arbitrary, but is related to the target rating of the bank. The risk
weighted asset is then found by further multiplying by terms such as
the exposure at default (i.e. size of the loan). Thus the capital
required for firm $i$ can be viewed as the loss at the 99.9th
percentile level of stress, in {\em excess} of the expected loss,
multiplied by a conversion factor. From this structure it is clear
that a key ingredient for the capital adequacy requirements is a
good model of credit risk that works well into the tail of the loss
frequency distribution.

Returning to our description of {\em default dynamics}, let us first
focus on the case of independent firms, with $J_{ij} = 0\ \forall
i,j$, and consider a single epoch for our model with fluctuating
forces given by (\ref{eq:noise_dec_gen}) at given macro-economic
condition $\eta_0$. The probability of a default of firm $i$ with
average unconditional monthly default probability $p_i$ occurring
during the epoch $t\to t+1$  in our model is given by
\begin{eqnarray}
\bra n_{i,t+1}|n_{i,t} = 0  \ket = \Phi\left( \frac{\sqrt{\rho_i}\,\eta_0
+\Phi^{-1}(p_i) }{\sqrt{1 -\rho_i}}\right)
\end{eqnarray}
Since the probability of default is increasing with $\eta_0$, we can
find the probability of default not exceeded at e.g. the 99.9
percent confidence level; it is given by setting $\eta_0 =
\Phi^{-1}(0.999)$ in the above equation (recall $\eta_0$ is
distributed as a zero-mean, unit-variance Gaussian random number).
As above, the excess capital required is the loss at the 99.9th
percent level minus the expected loss (multiplied by a risk factor).
However, when we consider the case of an interacting economy with
non-zero $J_{ij}$, we  find that in fact
\begin{eqnarray}
\bra n_{i,t+1}|n_{i,t} = 0 \ket = \Phi\left(\frac{J_0 m_t +\sqrt{\rho_i}\,\eta_0 -
\vartheta_i}{\sqrt{1 -  \rho_i + J^2 m_t}}\right)\ ,
\end{eqnarray}
where
\begin{eqnarray}
m_t = \frac1N \sum_j n_{j,t}
\end{eqnarray}
is the fraction of firms within the economy that have defaulted up to time $t$; we
also expressed the expected monthly default rate $p_i$ in terms of a `rescaled initial
wealth' $\vartheta_i$,  $\Phi^{-1}(p_i)= -\vartheta_i$.

Thus we find that our formulation is very similar to that used in
BASEL II. However, we directly take account of the correlations in
defaults caused by credit contagion. This introduces two extra
parameters into the model but it does markedly change the behaviour
in the tails of the loss frequency distribution, and thereby in the
tails of the loss distribution itself. Correlation between firms is
essentially a dynamic phenomenon --- if there is no dynamics, there is
no way for one's firms' performance to influence the viability of any
other firm. Thus rather than considering firms to be independent over
a single epoch which lasts the entire period of any loan, we split
the overall time (e.g. one year) into smaller units (e.g. one month)
and let the firms evolve over these smaller time units with the default
probability adjusted (since the default event in 12 monthly epochs is
compounded 12 times as opposed to a single epoch). A firm may
default at any point, but will then influence its partner firms for
the remainder of the time horizon. The complexity of the theory is
merely linear in time, thus it is not a great computational burden
to choose this approach.

Following the approach described in \cite{HatchettKuehn06} it is
possible to solve the model in a stochastic manner. Credit contagion
within this model is encoded at each time by a single number, the
fraction of firms that have defaulted thus far, which evolves according
to
\begin{eqnarray}
m_{t+1} = m_t + \left\langle \Big(1 - \bra n_t(\vartheta) \ket\Big)
\Phi\left (\frac{J_0 m_t +
\sqrt{\rho(\vartheta)}\,\eta_0 - \vartheta}{\sqrt{1 - \rho(\vartheta) + J^2 m_t}}
\right)\right\rangle_\vartheta
\label{eq:macro_dyn}
\end{eqnarray}
where $\bra n_t(\vartheta) \ket$  denotes the time-dependent monthly default
rate of firms with $\vartheta_i\approx \vartheta$, as influenced by interactions
with the economy (see Eq (16) of \cite{HatchettKuehn06}), and the larger angled
brackets with subscript $\vartheta$ denote an average over the bare monthly
probabilities of default for the ensemble of firms, or equivalently over the
distribution $p(\vartheta)$ of their rescaled initial wealth parameters
$\vartheta$. In (\ref{eq:macro_dyn}) the Basel II recommendation
which links correlations to macro-economic factors with (unconditional)
default probabilities, $\rho_i=\rho(p_i) \to \rho(\vartheta_i)$, via
(\ref{eq:Basel_corr}) is formally taken into account.

Note that credit contagion affects the dynamics of defaults only via two
parameters, $J_0$ and $J$, which characterise the mean and variance of the impact
parameter distribution. Note also that the parameter $\alpha$ which quantifies
forward-backward correlation of mutual impacts according to (\ref{eq:xij_stats})
does not appear in the final formulation, nor are there any memory-effects in
the dynamics, as would normally be expected for systems of this type. The reason
for this simplifying feature is in the fact that the defaulted state is taken
to be absorbing over the risk horizon of one year.

\section{Results}

We now turn to presenting a few key results of our analysis. Our
results concerning default dynamics and loss distributions are
obtained for an economy in which the parameters $\vartheta_i$
determining unconditional monthly default probabilities $p_i$
according to $\vartheta_i = - \Phi^{-1}(p_i)$,  are normally
distributed with mean $\vartheta_0 = 2.75$, and variance
$\sigma_\vartheta^2 = 0.1$ so that typical monthly bare default
probabilities are in the $10^{-5} - 10^{-3}$ range. The couplings to
the macro-economic factor are chosen to depend on the expected
default probabilities according to the Basel II prescription
(\ref{eq:Basel_corr}).

In Fig. 1 we we show that renormalisation (with respect to credit
contagion) makes little difference to the typical default dynamics.
The evolution of the fraction of defaulted firms in interacting
economies differs hardly from that of the non-interacting economy
with $J_{ij}=0 \Leftrightarrow (J_0,J)= (0,0)$.

\begin{figure}[h]
\begin{center}
\epsfig{file=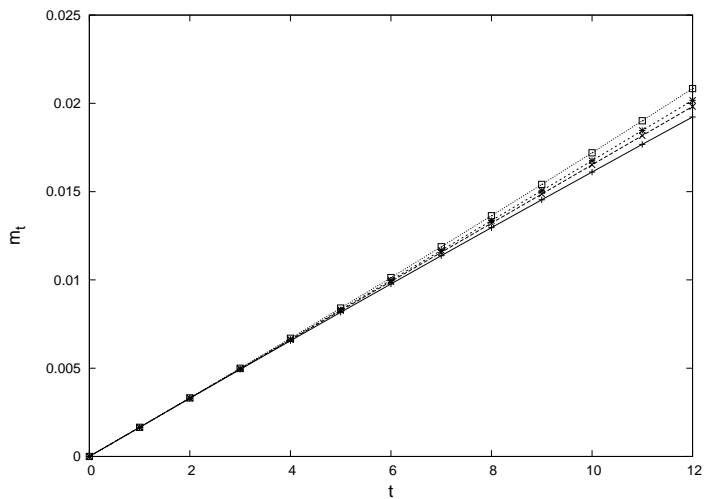,width=9.5cm}
\end{center}
\caption[]{Typical fraction of defaulted companies as a function of time
for $(J_0,J)= (0,0),(1,0),(0,1)$, and (1,1) (bottom to top), realized for
a neutral macro-economic factor $\eta_0=0$.}
\end{figure}

In marked contrast to this, the tails of loss-frequency and loss distributions
are strongly affected by the presence of interactions in the system, as shown
in Figs. 2 and 3. We note that the tails of the loss-frequency distribution and
the loss distribution are more pronounced than in our previous study
\cite{HatchettKuehn06}. This solely due to the fact that in the present paper
we followed the Basel II suggestion that relates the coupling of a company to
macro-economic factors with its default probability via (\ref{eq:Basel_corr}).

In computing the economy-wide losses per node for given
macro-economic condition $\eta_0$, \be L(\eta_0) = \frac{1}{N}
\sum_i n_{i,12} \ell_i \label{Leta0} \ee we assume that the $\ell_i$
are randomly sampled from the loss distribution for node $i$, taken
to be independent of the stochastic evolution, but correlated with
the bare monthly default probability. In the large $N$ limit this
gives \be L(\eta_0)= \lim_{N\to\infty} \frac{1}{N}\sum_i
n_{12}(\vartheta_i) \ell_i = \int d\vartheta p(\vartheta) \langle
n_{12}(\vartheta)\rangle \overline \ell(\vartheta) \ee by the law of
large numbers, where $\overline \ell=\overline \ell(\vartheta)$ is
the mean of the loss distribution for a node with default
probability $p_d(\vartheta)$. As an example we consider an economy
where average losses are inversely proportional to the unconditional
default probabilities $p_i = p_d(\vartheta_i)= \Phi(-\vartheta_i)$,
\be \overline \ell(\vartheta) =\frac{\ell_0}{\varepsilon +
p_d(\vartheta)} \label{scalelbar} \ee with a parameter $\varepsilon$
preventing divergence as $p_i\to 0$. In this way, the contribution
to the total losses will be approximately uniform over the bands of
different default probabilities. The distribution of the
economy-wide losses per node is driven by the distribution of the
macro-economic factor, and is computed analytically as shown in
\cite{HatchettKuehn06}. A typical result is shown in Fig. 3, for
which we chose the scale $\ell_0=1$ and the regularizer
$\varepsilon= 0.005$. Once more economic interactions are seen to
strongly affect the tail of the loss distribution at large losses,
which is due to the possibility of avalanches of loss events in
times of extreme economic stress.

\begin{figure}[h]
\begin{center}
\epsfig{file=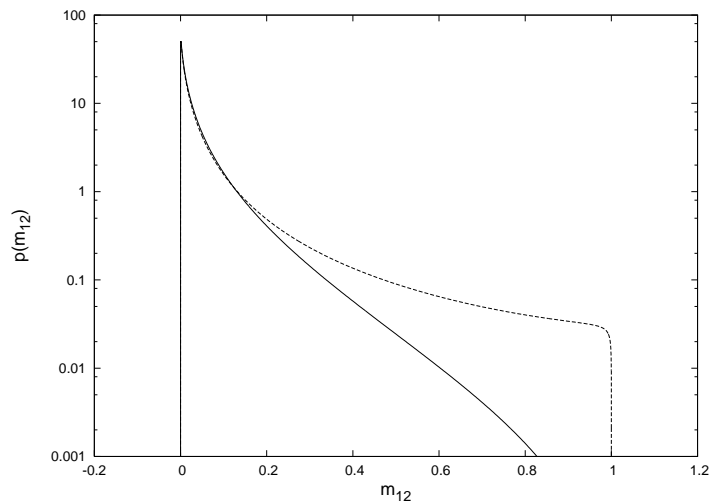,width=9.5cm}
\end{center}
\caption[]{Distribution of the fraction of defaulted companies
for $(J_0,J)= (0,0)$ (bottom) and $(J_0,J)=(1,1)$ (top).}
\end{figure}

\begin{figure}[h]
\begin{center}
\epsfig{file=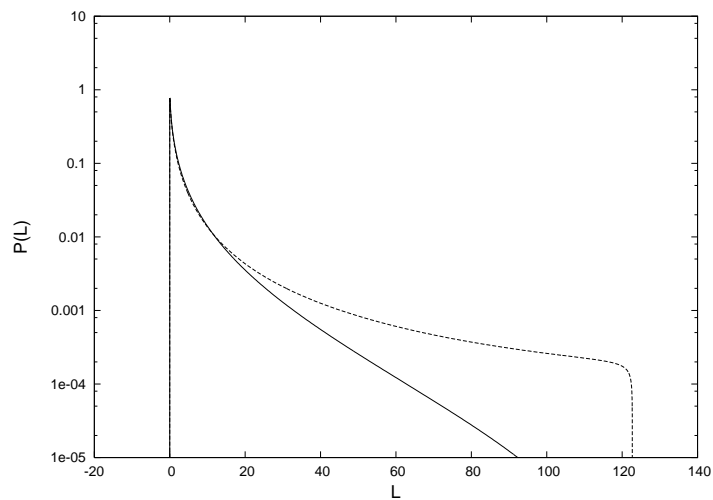,width=9.5cm}
\end{center}
\caption[]{Distribution of losses for the non-interacting system with
$(J_0,J)= (0,0)$ (bottom) and the interacting economy with $(J_0,J)=(1,1)$ (top).}
\end{figure}

Note that we have been dealing here with ``synthetic" parameter
distributions for averages of loss distributions, as well as bare
monthly failure probabilities. These could be replaced by realistic
ones without affecting the general set-up. We have not looked
specifically at finite size effects here. In \cite{HatchettKuehn06}
it was shown that they are fairly small.

\section{Conclusion}
In this paper we have looked to incorporate the risk due to credit
contagion into the internal ratings based approach discussed in
BASEL II. While the mathematical subtleties are discussed in full
detail elsewhere \cite{HatchettKuehn06}, essentially the large
number of neighbours assumed for firms means that the law of large
numbers and central limit theorems apply to the interactions,
meaning that our theory requires only two more parameters than the
BASEL II approach. In terms of risk, one of the striking results is
that while the effect of interactions is relatively weak in typical
economic scenarios, it is pronounced in times of large economic
stress, which leads to a significant fattening of the tails of the
portfolio loss distribution. This has implications on the fitting of
loss distributions to historical data, where care must be taken not
only to fit the average behaviour but also to take care with the
more extreme events.

This touches the issue of model calibration  discussed in greater
detail in  \cite{NeuKuehn04}. We note that our model requires bare
default probabilities and conditional default probabilities as
inputs. Historical data, however only contain
interaction-renormalised default probabilities, and thus the problem
arises of how to disentangle the two effects. Concerning typical
behaviour, Fig. 1 shows that the effect of interactions is fairly
small, and interaction-renormalised default probabilities can, to a
first approximation within this model, be taken as substitutes for
the bare ones. Concerning conditional default probabilities, these
would have to be obtained from refined rating procedures; see
\cite{NeuKuehn04}. Interestingly, however, only the low order
statistics of these are needed to describe the collective dynamics
of the system. Their effect manifests itself only in situations of
economic stress, generating fat tails in portfolio loss
distributions.

The model we have proposed is relatively simple in two important respects.
Firstly, we do not take into account credit quality migration but have
just two states for our firms, solvent or defaulted. The model could be
extended to allow for more states for each firm, although the full
complexity of non-Markovian dynamics would resurface in an attempt to
take credit quality migration along these lines into account.
Secondly, the firms and their environment are rather homogeneous, which
in practical situations is of course an approximation. This approximation
has been made for convenience rather than out of necessity; the techniques
described in \cite{HatchettKuehn06} can be adapted so as to treat situations
with more heterogeneity in local environments.
We intend to work on some of these possible model generalisations in the future.

One advantage of our simple model is that it is exactly solvable and the
solution itself is not overly involved theoretically or computationally, and
we only need to introduce two extra parameters to quantify the effect of economic
interactions --- compared to the BASEL II approach, which ignores credit
contagion altogether.

\paragraph{Acknowledgements} We thank Peter Neu for useful discussions and
helpful remarks. This paper has been prepared by the authors as a
commentary on the topic as at September 2006 and is not a definitive
analysis of the subject matter covered.  It does not constitute
advice or guidance in this area and should not be relied upon as
such.  The paper has not been prepared by J.H. in his capacity as an
employee of Hymans Robertson LLP and the views expressed do not
represent those of Hymans Robertson LLP. Neither the authors nor
Hymans Robertson LLP accept liability for any errors or omissions

\section*{References}

\end{document}